\documentclass[aps]{revtex4}

\begin{document}
\draft

%
%  Uncomment following two lines and one below for 2 column format.
%@twocolumnfalse\endcsname
%\tightenlines

\preprint{Nisho-07/2}
\title{Thermalization of Color Gauge Fields \\
in High Energy Heavy Ion
Collisions}
\author{Aiichi Iwazaki} 
\address{International Politics and Economics, Nishogakusha University, \\
Ohi Kashiwa Chiba
  277-8585,\ Japan.} 
\date{Dec. 10, 2007}
\begin{abstract}
We discuss quantum mechanical decay of color magnetic field generated initially
at high energy heavy ion collisions. It is caused by Nielsen-Olesen unstable modes
and is accomplished possibly in a period $<1$fm/c.
We show that the decay products, i.e. incoherent gluons
may be thermalized in a sufficiently short period ($< 1$fm/c ).
The precise determination of the period is made by
calculating two point function of the color magnetic field 
in a model of color glass condensate.
\end{abstract}	
\hspace*{0.3cm}
\hspace*{1cm}

\maketitle
%%%%%%%%%%%%%%%%%%%%%%%%%%%%%%%%%%%%%%%%%%%%%%%%%
%\tightenlines

In high energy collisions of heavy ions in RHIC or LHC, the most important
ingredients for producing dense quark gluon plasma ( QGP ) are small x gluons in the nuclei~\cite{1,2,3,4,5}.
The gluons with small transverse momenta are sufficiently dense in the nuclei and so
they may be treated as classical fields produced by glassy large x gluons
even after the collisions of the nuclei. It has been shown that 
longitudinal color magnetic and electric fields of the small x gluons are generated 
initially at the collisions. They are classical fields and evolve classically
according to a model of color glass condensate (CGC)\cite{cgc}.
It is expected phenomenologically\cite{hirano} 
that the decay of the gauge fields within a period of $1$fm/c after the collisions leads to thermalized QGP.

In an extremely high energy collision, the radius of nuclei is nearly zero due to the Lorentz contraction.  
Thus, the initial gauge fields have only transverse momentum perpendicular to the collision axis, but
have no longitudinal momentum ( rapidity ). Such classical gauge fields can not
possess any longitudinal momentum in their classical evolution, since equations of motion of the gauge fields
are invariant under the Lorentz boost along the collision axis\cite{2,3,4}.
It has recently been shown\cite{ve} that the addition of small
fluctuations with rapidity to the initial gauge field
induces exponentially increasing modes with longitudinal momentum. Such fluctuations may be
produced by quantum fluctuations around the classical gauge fields\cite{fuku,review}.
The production of the exponentially increasing modes implies that a process toward
thermalization has started; the decay of the gauge field and the isotropization
of momenta.
Although the modes increase exponentially in time, 
the time needed for them to grow sufficiently large is 
too long to be 
consistent with the phenomenological expectation. 

In this letter we show that the decay of the color magnetic field is 
caused quantum mechanically by Nielsen-Olesen unstable modes\cite{no,savvidy}.
It can occur in a sufficiently short period after the collisions so that
the phenomenological expectation is satisfied. 
The decay of the gauge field produces much dense incoherent gluons. 
They carry longitudinal momenta as well as transverse momenta; their typical momenta are 
given by the square root, $\sqrt{g\bar{B}}$, 
of a color magnetic field, $g\bar{B}$, averaged over transverse plane.
It is shown with reasonable choice of parameters that
such dense gluons have sufficiently short mean free path
to be thermalized immediately after their productions.
Consequently, we obtain isotropic thermalized QGP.
The life time of the gauge field 
is given 
in terms of a saturation momentum used in a model of CGC, e.g.
the Mcleran-Venugopalan ( MV ) model\cite{1}. It can be obtained by
the calculation of a two point function of color magnetic fields 
in arbitral proper time, $\tau\equiv \sqrt{t^2-x_3^2}>0$ when
the collisions occur at $\tau=0$. Here, without explicitly performing the calculation,
we simply give general formulae for the quantities such as number density of produced gluons,
their mean free path and temperature of QGP.
All of them are expressed in terms of a single parameter, $Q_s\tau_c$;
$Q_s$ is saturation momentum used in the MV model and $\tau_c$ is the quantum mechanical life time of the magnetic field.
Thus, once we obtain the value of $Q_s\tau_c$, 
we can determine whether or not the thermaliztion of QGP is achieved at $\tau=\tau_c$.
Indeed, when we take phenomenologically reasonable values,
$Q_s=2$GeV and $\tau_c=0.5$fm/c, or $Q_s\tau_c \geq 5$,
thermalized QGP is realized at $\tau=\tau_c$.
In this paper we only discuss the
fate of the color magnetic field and assume that the color electric field
is screened immediately by dense small x gluons.

First, we sketch briefly Nielsen-Olesen unstable modes generated under the color magnetic field
and the subsequent decay of the field using SU(2) gauge theory.
The modes arise in the presence of a homogeneous color magnetic field, $B^a_i$. Without loss of generality,
we may point the field into the direction of 
third axis both in color space, $a$, and real space, $i$; $B^a_i=B \delta^{a,3}\delta_{i,3}$.
Then, we decompose the
Lagrangian of gluons
with the use of the variables, "electromagnetic field" 
$A_{\mu}=A_{\mu}^3,\,\,\mbox{and} \,\, \mbox{"charged vector field"}\,
\Phi_{\mu}=(A_{\mu}^1+iA_{\mu}^2)/\sqrt{2}$ 
where indices $1\sim 3$ denote color components,

\begin{eqnarray}
\label{L}
L&=&-\frac{1}{4}\vec{F}_{\mu
  \nu}^2=-\frac{1}{4}(\partial_{\mu}A_{\nu}-\partial_{\nu}A_{\mu})^2-
\frac{1}{2}|D_{\mu}\Phi_{\nu}-D_{\nu}\Phi_{\mu}|^2- \nonumber \\
&+&ig(\partial_{\mu}A_{\nu}-\partial_{\nu}A_{\mu})\Phi_{\mu}^{\dagger}\Phi_{\nu}+\frac{g^2}{4}(\Phi_{\mu}\Phi_{\nu}^{\dagger}-
\Phi_{\nu}\Phi_{\mu}^{\dagger})^2
\end{eqnarray}
with $D_{\mu}=\partial_{\mu}+igA_{\mu}$,
where we have used a gauge condition, $D_{\mu}\Phi_{\mu}=0$. 
We note that the third term in the right hand side represents anomalous magnetic moment of the field, $\Phi_{\mu}$.

Using the Lagangian we find that the energy $E$ of
the charged vector field $\Phi_{\mu}\propto e^{-iEt}$ 
in the magnetic field, $B=\epsilon_{i,j}\partial_iA_j$, is given by
$E^2=k_3^2+2gB(n+1/2)\pm 2gB$.
Here,
$\pm 2gB$ ( the integer $n\geq 0$ ) denote contributions of spin components
of $\Phi_{\mu}$ ( Landau levels )
and $k_3$ denotes 
momentum parallel to the magnetic field.
The term of $\pm 2gB$ represents the contribution of the anomalous magnetic moments. 

Obviously, the modes with $E^2(n=0,k_3^2<gB)<0$ are unstable;
their amplitudes increase ( decrease ) exponentially in time.
We call them Nielsen-Olesen unstable modes.
Their presence implies that the color magnetic field is unstable;
it decays with the production of these unstable gluons.
The decay width, $\Gamma$, in unit volume has been calculated such as
$\Gamma=\frac{(gB)^2}{8\pi}$ by assuming a homogeneous color magnetic field, $B$.

We can show\cite{meeting} that main results on the classical evolution of gauge fields   
in the Ref.\cite{ve} are understood in terms of these Nielsen-Olesen unstable modes. Namely,
the exponentially increasing modes found in the Ref.\cite{ve} can be
identified as Nielsen-Olesen unstable modes. Actually,
by solving linearlized equation of motions\cite{fu} for 
the unstable modes under both homogeneous longitudinal
color magnetic and electric field taken appropriately,
we can find the followings;
$1)$ maximum longitudinal momentum, $\nu_{\rm{max}}$, excited increases linearly
with time, $\tau$,  $2)$ longitudinal momentum, $\nu$, giving the maximum Fourier component of longitudinal pressure
is small and increases very slowly with $\tau$, and $3)$ the maximum Fourier component of the pressure
increases in such a way as $\exp(\tau^{3/4})$, although it increases as $\exp(\tau^{1/2})$ in the Ref.\cite{ve}.
( Numerical results in the Ref.\cite{ve} can be fitted even by using this increasing function $\exp(\tau^{3/4})$. )
The saturation of the exponentially increasing component of the pressure arises due to the quartic interactions
of $\Phi_{\mu}$ in our analysis. As has been recognized, the isotropization
of momenta in the classical evolution needs too much time, $\tau_{\rm{iso}}$, to be consistent with 
the phenomenological expectation, $\tau_{\rm{iso}}<1$ fm/c.  
Therefore,
it is natural to see their quantum effects by using the above
decay formula. All of the unstable modes contribute to the decay quantum mechanically.

The unstable gluons of the exponentially increasing modes occupy the lowest Landau level ( $n=0$ ). Their wave functions
are characterized by angular momentum, $m=0,1,2,,,$ around $x_3$ axis,

\begin{equation}
\Phi(z,x_3,t;m,k_3)\equiv(\Phi_1-i\Phi_2)\sqrt{l/2}=g_m z^m \exp(ik_3x_3-|z|^2/4l^2+t\sqrt{gB-k_3})
\end{equation}
with $z=x_1+ix_2$ and normalization factor, $g^2_m\equiv \frac{1}{\pi m!(2l^2)^{m+1}}$.
Here we chosen the central gauge, $A_i=(-Bx_2/2,Bx_1/2,0)$.
In this formula $l\equiv 1/\sqrt{gB}$ denotes magnetic length representing
transverse extension of the states. 
The factor of $\exp(t\sqrt{gB-k_3^2})$
represents relative production ratio of the modes with each longitudinal momentum, $k_3$.
All of the states specified by $m$ are degenerate in energy. 
It has been shown\cite{screen,screen2} that the states screen the original magnetic field in a way just like
a Landau diamagnetism. 
Hence, the color magnetic field decays with the production 
of the Nielsen-Olesen unstable modes. This is very similar to the case
that an electric field decays with the production of charged particles,
which screen the electric field.

It is obvious that both transverse and longitudinal
momenta of the produced gluons are typically given by $1/l=\sqrt{gB}$. 
Although the color magnetic field
is homogeneous in the longitudinal direction, 
the gluons have the nontrivial longitudinal momentum.
This is sharply contrasted to the case of classical evolution of the gauge fields. 
In the classical evolution, longitudinal momentum never arises dynamically
due to the symmetry of initial condition and equation of motion of gauge fields. 

The collisions between a nuclei running into $+x_3$ direction and a nuclei running into $-x_3$ direction
are supposed to occur at $t=x_3=0$. In the high energy collisions of identical heavy ions, 
a longitudinal gauge field of small x gluons is generated initially at
the collisions.
The explicit form of the gauge field at $\tau=\sqrt{t^2-x_3^2}=2\sqrt{x^+x^-}=0$ is given 
in the MV as,

\begin{equation}
A_i(\tau=0,x_t)=a_{1,i}(x_t)+a_{2,i}(x_t) \quad \mbox{and} \quad
a_{s,i}(x_t)=\frac{i}{g}U_s(x_t)\partial_iU^{\dagger}_s(x_t),
\end{equation}
where gauge potentials, $a_{s,i}(x_t)$ are produced by the sources, $\rho_s(x_t,z^{\pm})$ of large x gluons in s-th nuclei,
$U_1(x_t)=P\exp(-ig\int^{x^-}dz^-\frac{1}{\partial^2}\rho_1(x_t,z^-)\delta(x^-))$ and
$U_2(x_t)=P\exp(-ig\int^{x^+}dz^+\frac{1}{\partial^2}\rho_2(x_t,z^+)\delta(x^+))$ where
$\partial^2$ denotes derivative in transverse coordinates, $x_t=(x_1,x_2)$.
The distribution of the sources is given such that
$\langle\rho_s(x_t,x^{\pm})\rho_{s'}(y_t,y^{\pm})\rangle
=\delta_{s,s'}Q_s^2/g^2\delta^2(x_t-y_t)\delta(x^{\pm}-y^{\pm})$.
The gauge potentials represents longitudinal color magnetic field, $B_3(x_t)=ig\epsilon_{ij}[a_{1,i}(x_t),a_{2,j}(x_t)]$.
Since this magnetic field has typically a transverse momentum,
$Q_s$, it is never homogeneous in transverse directions. 

Although the color magnetic field is not homogeneous in space and time, it may decays due to 
the excitation of Nielsen-Olesen unstable modes mentioned above. 
We need to find an effective homogeneous magnetic field
relevant to the decay in order to use the formula of the decay width, $\Gamma$, in the above.
It is reasonable to suppose that
such a field is given by a color magnetic field averaged
over the transverse space. In order to obtain the average field explicitly, 
we use root mean square of color magnetic flux, $\hat{\Omega}(\tau)\equiv \int_S d^2x_t B_3(x_t,\tau)$,

\begin{equation}
g\bar{B}(\tau)\equiv g\frac{\sqrt{\Omega^2(\tau)}}{S} \quad 
\mbox{with}\quad \Omega^2(\tau)=\rm{Tr}\langle \hat{\Omega}(\tau)\hat{\Omega}(\tau)\rangle
\quad \mbox{and}\quad \langle \hat{\Omega}(\tau) \rangle=0,
\end{equation}
where the average, $\langle \hat{Q} \rangle$ is taken over the sources, $\rho_s$.
The integral, $\int_Sd^2x_t$, is taken over a transverse region with 
collision area, $S$, relevant to a heavy ion collision, e.g. $S=\pi R_N^2$ with nuclear radius, $R_N$,
for central collisions.
Since the correlation length of the magnetic field is given roughly by $Q_s^{-1}$, we may rewrite $\Omega^2$ such as
$\Omega^2=\frac{c(\tau)Q_s^2S}{g^2}$ where the numerical coefficient $c(\tau)$ can be obtained by numerical
calculation of the correlation function, Tr$\langle B_3(x_t,\tau)B_3(y_t,\tau)\rangle$ in the formula.
Thus, the homogeneous average magnetic field can be rewritten as
$g\bar{B}=Q_s\sqrt{c(\tau)/S}$. 
Since the correlation function, Tr$\langle B_3(x_t,\tau)B_3(x_t,\tau)\rangle$,
has been shown\cite{5} to decrease as $1/\tau$ for $Q_s\tau \geq 1$, we may speculate
the similar decrease of Tr$\langle B_3(x_t,\tau)B_3(y_t,\tau)\rangle \sim 1/\tau$. 
It means that $c(\tau)\sim 1/Q_s\tau$. ( It is easy to see that perturbative solutions, $B_3(x_t,\tau)$, of
gauge fields behaves such as $B_3(x_t,\tau)\propto 1/\sqrt{\tau}$. Thus, 
Tr$\langle B_3(x_t,\tau)B_3(y_t,\tau)\rangle \sim 1/\tau$ for $Q_s\tau \geq 1$. )
Note that since $c(\tau)$ does not involve a parameter with mass dimension except for $Q_s$,
it depends on $\tau$ only through the combination, $Q_s\tau$.

We assume this field as an effective homogeneous color magnetic field
generating the unstable Nielsen-Olesen modes. 
Actual decay of the inhomogeneous magnetic field
is more complicated than the simple case discussed in the present paper.
But, by using the effective magnetic field 
we may obtain approximately the life time of the field
and the number density of gluons produced subsequently.
In order to determine more accurately these quantities,
we need to examine more closely how the 
inhomogeneous color magnetic field decays.

We should make a comment on a consistency condition that Nielsen-Olesen unstable modes 
have to be located inside of
the region involving the magnetic field. This constraints the parameter, $g\bar{B}$,
such as $1/\sqrt{g\bar{B}}< \sqrt{S/\pi}$. Namely, the magnetic length, $1/\sqrt{g\bar{B}}$, is smaller than the
radius, $\sqrt{S/\pi}$, of the collision area. This condition is satisfied in general even for
peripheral collisions as far as $Q_s^2S > \pi^2/c(\tau)$ with $Q_s>1$ GeV.

As we can see soon below, the decay of the magnetic field is very rapid.
On the contrary, the dependence of $g\bar{B}$ on $\tau$ is much smooth
; $g\bar{B}=Q_s\sqrt{c(\tau)/S}\propto \tau^{-1/2}$. Hence,
we may apply the above decay formula to $g\bar{B}$ depending on $\tau$. 
Consequently,
we find the life time of the magnetic field by solving the following equation,

\begin{equation}
\label{gamma}
1=\Gamma SR_z\tau_c=\frac{(g\bar{B})^2S\tau_c^2}{8\pi}=\frac{c(\tau_c)Q_s^2\tau_c^2}{8\pi},
\end{equation}
where $R_z=\tau$ denotes longitudinal extension of the color magnetic field.
( In the explanation of Nielsen-Olesen unstable modes we have used standard time coordinate, $t$, not 
$\tau=\sqrt{t^2-x_3^2}$. Thus, we can apply the decay formula of $B$ only to midrapidity, 
$\eta=\log(\sqrt{\frac{t+x_3}{t-x_3}})\simeq 0$. )
It is important to note that $\tau_c$ does not depend on 
transverse area, $S$. It implies that the life time of magnetic field does not depend
on the centrality of the collisions.

Solving the equation(\ref{gamma}), we obtain the value of $Q_s\tau_c$.
Thus, if $Q_s$ becomes larger, the life time, $\tau_c$, of 
the color magnetic field becomes smaller. Since the saturation momentum, $Q_s$, is 
larger in LHC than that in RHIC,
QGP is expected to be thermalized
much earlier in LHC that that in RHIC.

Although the precise evaluation of $c(\tau)$ is needed for solving the equation,
we may estimate roughly $\tau_c$ by assuming a naive correlation, Tr$\langle B_3(x_t)B_3(y_t)\rangle
=Q_s^4/g^2$ for $|x_t-y_t|<Q_s^{-1}$ and vanishes for otherwise. Then, it leads to $\Omega^2\simeq Q_s^2S/g^2$,
namely, $c=1$. Hence, we find $\tau_c\simeq 5 Q_s^{-1}\simeq 0.5\,\mbox{fm/c}$ for $Q_s=2$GeV.
This satisfies the phenomenological requirement, $\tau_c<1$fm/c. Although this estimation
is very rough, the precise evaluation of $c(\tau_c)$ would similarly give 
a phenomenologically reasonable value in the following way.

Since the real correlation, Tr$\langle B_3(x_t)B_3(y_t)\rangle$, has no such sharp
boundary like $\exp(-Q_sr)$ in $r\equiv |x_t-y_t|$ as assumed above,
the integral $\int d^2x_t$ gives much larger contribution than one estimated just above.
This results in $c(\tau=0)\gg1$. On the other hand $c(\tau)$ decays as $c(\tau)\propto 1/Q_s\tau$.  
These would lead to $c(\tau_c)\sim O(1)$. Thus, we expect that
an appropriate result, e.g. $\tau_c=0.5$ fm/c, would be also obtained 
even with the precise evaluation of
Tr$\langle B_3(x_t)B_3(y_t)\rangle$. 

We now proceed to show that gluons produced by the decay of the magnetic field 
are sufficiently dense to be thermalized quickly 
when a phenomenological requirement, $\tau_c\leq 1$ fm/c, is satisfied with $1\,\mbox{GeV}<Q_s<3$ GeV. 
We give a general formula for the density of gluons and determine whether or not 
the gluons are thermalized. The thermalization is assumed to be achieved when a produced gluon
interacts with others many times within a period, $\tau_c$. We also give 
their temperature when the gluons are thermalized. The formulae are expressed in terms of
the parameter, $Q_s\tau_c$.

The momentum distribution of the gluons produced by the decay is proportional to absolute square of the wave function
of the unstable gluons, that is, $|\Phi(k_t,k_3,t;m)|^2=|\int d^2x_t\exp(-ik_tx_t)\Phi(x_t,x_3,t;m,k_3)|^2$.

Hence,
the gluon number density is given by 

\begin{equation}
\label{N}
N(\tau)=N_0(\tau)\int d^2k_t dk_3 \sum_{0\leq m\leq N_R}|\Phi(k_t,k_3,\tau;m)|^2,
\end{equation}
with normalization constant, $N_0(\tau)$,
where we have summed equally over the wave functions with different $m$
since they are degenerate.
The largest angular momentum, $N_R$ is determined by the condition that
the gluons must be located inside the transverse region with surface area of $S$:
Because the average $\langle\Phi||z|^2|\Phi\rangle=2(m+1)l^2$ should be less than $S/\pi$,
$N_R=\frac{1}{2}(R/l)^2-1\simeq \frac{1}{2}(R/l)^2=\sqrt{2SQ_s^2}/Q_s\tau$ with $\pi R^2=S$. On the other hand,
the normalization constant, $N_0(\tau_c)$, is determined in the following.
Namely,
the energy density of the produced gluons
is equal to the energy density of the color magnetic field, 
$B^2(\tau)/2\equiv \rm{Tr}(\langle B(x_t,\tau)B(x_t,\tau)\rangle)/2$, 

\begin{equation}
\frac{(gB(\tau_c))^2}{2g^2}=N_0(\tau_c)\int d^2k_tdk_3 \sqrt{k_t^2+k_3^2}\sum_{0\leq m\leq N_R}|\Phi(k_t,k_3,\tau_c;m)|^2,
\end{equation} 
where we have used the energy, $\sqrt{k_t^2+k_3^2}$, of free gluons 
because the gluons become almost free after the decay of the magnetic field. 
We have also assumed the translational invariance in the transverse directions. 
Thus, $N_0(\tau_c)$ is given by

\begin{equation}
N_0(\tau_c)=\frac{(gB(\tau_c))^2}{2g^2\int d^2k_tdk_3 \sqrt{k_t^2+k_3^2}\sum_{0\leq m\leq N_R}|\Phi(k_t,k_3,\tau_c;m)|^2}
\quad .
\end{equation}
Inserting the formula into eq(\ref{N}), we find the number density of the gluons,

\begin{equation}
N(\tau_c)=\frac{(gB(\tau_c))^2}{2g^2}\frac{\int d^2k_t dk_3 \sum_{0\leq m\leq N_R}|\Phi(k_t,k_3,\tau_c;m)|^2}
{\int d^2k_tdk_3 \sqrt{k_t^2+k_3^2}\sum_{0\leq m\leq N_R}|\Phi(k_t,k_3,\tau_c;m)|^2}\quad .
\end{equation}
Noting that

\begin{eqnarray}
|\Phi(k_t,k_3,\tau;m)|^2&=&|4l^2\pi g_m(2il^2)^m(k_1+ik_2)^m\exp(ik_3x_3-k_t^2l^2+\tau\sqrt{g\bar{B}-k_3^2})|^2 \\
&=&8l^2\pi \frac{(2l^2k_t^2)^m}{m!}\exp(-2l^2k_t^2+2\tau\sqrt{g\bar{B}-k_3^2})\\
&=&8l^2\pi \frac{(2\sigma_t^2)^m}{m!}\exp(-2\sigma_t^2+2\bar{\tau}\sqrt{1-\sigma_3^2})
\end{eqnarray}
with $l^2=1/g\bar{B}(\tau)$, $\sigma_t\equiv lk_t$, $\sigma_3\equiv lk_3$ and $\bar{\tau}\equiv \tau/l$,
we obtain

\begin{eqnarray}
N(\tau_c)&=&\frac{1}{2g^2l^3}\frac{(gB(\tau_c))^2}{(g\bar{B}(\tau_c))^2}\frac{\int_{|\sigma_3|\leq 1} d^2\sigma_t d\sigma_3
\sum_{m=0}^{N_R}\frac{(2\sigma_t^2)^m}{m!}\exp(-2\sigma_t^2+2\bar{\tau}_c\sqrt{1-\sigma_3^2})}
{\int_{|\sigma_3|\leq 1} d^2\sigma_t d\sigma_3 \sqrt{\sigma_t^2+\sigma_3^2}
\sum_{m=0}^{N_R}\frac{(2\sigma_t^2)^m}{m!}\exp(-2\sigma_t^2+2\bar{\tau}_c\sqrt{1-\sigma_3^2})}\\
&=&\frac{d(\tau_c)Q_s^3\sqrt{Q_s\tau_c}}{2g^2}\biggl(\frac{SQ_s^2}{8\pi}\biggr)^{1/4}\, f(N_R,\bar{\tau}_c)
\end{eqnarray}
with $(gB(\tau_c))^2\equiv d(\tau_c)Q_s^4$ and $\bar{\tau}_c=(8\pi/SQ_s^2)^{1/4}\sqrt{Q_s\tau_c}$,
where,

\begin{equation}
f(N_R,\bar{\tau}_c)\equiv \frac{\int_{|\sigma_3|\leq 1} d^2\sigma_t d\sigma_3
\sum_{m=0}^{N_R}\frac{(2\sigma_t^2)^m}{m!}\exp(-2\sigma_t^2+2\bar{\tau}_c\sqrt{1-\sigma_3^2})}
{\int_{|\sigma_3|\leq 1} d^2\sigma_t d\sigma_3 \sqrt{\sigma_t^2+\sigma_3^2}
\sum_{m=0}^{N_R}\frac{(2\sigma_t^2)^m}{m!}\exp(-2\sigma_t^2+2\bar{\tau}_c\sqrt{1-\sigma_3^2})}.
\end{equation} 
Here we note that the value of $d(\tau)\equiv(gB(\tau))^2/Q_s^4$ has already been calculated in the reference\cite{5}
such as $d(\tau)\simeq 0.15(1/Q_s\tau)$ for $Q_s\tau>1$. 

Up to now, we give a general formula of the gluon density in terms of 
a quantity, $Q_s\tau_c$ or $c(\tau_c)=8\pi/(Q_s\tau_c)^2$, 
which can be obtained by calculating the quantity,
$\int d^2x_td^2y_t\rm{Tr}\langle B_3(x_t,\tau)B_3(y_t,\tau)\rangle=\frac{c(\tau)Q_s^2S}{g^2}$ in the MV model.
We should note that
the parameters in $N(\tau_c)$, such as
$c(\tau_c)=8\pi/(Q_s\tau_c)^2$,
$N_R=\sqrt{2SQ_s^2}/Q_s\tau_c$ and $\bar{\tau}_c=(8\pi/SQ_s^2)^{1/4}\sqrt{Q_s\tau_c}$,
are all expressed in terms of $Q_s\tau_c$. 
Thus, once we determine the parameter, $Q_s\tau_c$ as well as $Q_s$ and $g^2$, 
we can obtain the number density of the gluons produced at $\tau=\tau_c$.

Although we have not yet determined $Q_s\tau_c$ by calculating explicitly $c(\tau)$, 
we will estimate 
how large is the number density, $N(\tau_c)$ 
and determine whether or not the gluons are thermalized.   
Assuming a phenomenologically expected value of $\tau_c=0.5$ fm/c with $Q_s=2$ GeV and $R=6.4$ fm,
we obtain $N(\tau_c)\sim 21/\mbox{fm}^3$ with $g=2$.
%$ SQ_s^2=360$,$N_R \simeq 30$, $\tau_c=0.5$ fm/c and $\bar{\tau}_c\simeq 1.1$.
Since the number density is very large,
the gluons may be thermalized quickly. Actually, the cross section, $\sigma$, of the gluons with the typical
momentum, $\sqrt{g\bar{B}}$ ( $=Q_s(\sqrt{Q_s\tau_c})^{-1}(8\pi/SQ_s^2)^{1/4}=188$ MeV), 
is approximately given by $\sigma=\alpha_g^2/g\bar{B}$
( $\alpha_g\equiv g^2/4\pi$ ).
Hence, the mean free path, $l_{\rm{m}}$, is that $l_{\rm{m}}=1/\sigma N(\tau_c)\simeq 0.06$ fm,
where we have used $\alpha_g=1$ since
the typical momentum of the gluons is $\sqrt{g\bar{B}}=188$ MeV.
Thus, the number of collisions 
for a gluon to interact with others within the period, $\tau_c=0.5$ fm/c,
is given such that $\tau_c \rm{c}/$$l_{\rm{m}}\simeq 8$ times.
This may be sufficiently large that the gluons are thermalized
within the period, $\tau_c=0.5$ fm/c. 
Therefore, after the decay of the color magnetic field,
much dense gluons are produced and are thermalized immediately.

We have assumed the parameter such as $Q_s\tau_c=5$ in the above estimation. 
If we assume a larger one, e.g. $Q_s\tau_c=10$ with $Q_s$ fixed, 
the number of collisions within a period, $\tau_c$ becomes larger and so
the thermalization proceeds more efficiently.
This is because the number of collisions behaves as $\tau_c\rm{c}/$$l_{\rm{m}}\propto \sqrt{Q_s\tau_c}$ since
$N(\tau_c)\propto 1/\sqrt{Q_s\tau_c}$ and $\sigma\propto Q_s\tau_c$.
While, when we increase the saturation momentum, $Q_s$ with $Q_s\tau_c$ fixed,
the number of collisions increases as $\tau_c\rm{c}/$$l_{\rm{m}} \propto Q_s^{3/2}$
since $N(\tau_c)\propto Q_s^{7/2}$ and $\sigma\propto Q_s^{-1}$. Hence,
the thermalization proceeds more efficiently in higher energy collisions of heavy ions.
Similarly, when we take larger collision area, $SQ_s^2$ than the previous case, 
the thermalization proceeds more efficiently since $\tau_c\rm{c}/$$l_{\rm{m}}$ ($\propto (SQ_s^2)^{3/4}$).
This implies that when the collision is central, the most efficient thermalization is 
achieved. ( The dependence of $f(N_R,\bar{\tau}_c)$ on $Q_s\tau_c$ is not so serious
to change significantly the above results. )

Finally, by taking the parameters used above such as $Q_s=2$ GeV, $\tau_c=0.5$ fm/c,
we estimate the temperature, $T_c$, of the QGP by assuming a gas of free quarks and gluons,

\begin{equation}
(\frac{7n_q}{8}+n_g)\frac{\pi^2}{30}T_c^4=\frac{(gB(\tau_c))^2}{2g^2}=d(\tau_c)Q_s^4/2g^2
\end{equation} 
with gluon's ( quark's ) number of degree of freedom, $n_g=16$ ( $n_q=36$ ). 
Since $d(Q_s\tau_c=5)\simeq 0.03$, we find that $T_c\simeq 246$ MeV.

To summarize, 
we have shown that the coherent color magnetic field generated initially in high energy heavy ion collisions
decays to produce incoherent dense gluons.
The decay occurs owing to the quantum effects of Nielsen-Olesen unstable modes.
The life time of the magnetic field has been given in
terms of the quantity, $c(\tau)$, which is obtained by
the evaluation of Tr$\langle B_3(x_t,\tau)B_3(y_t,\tau)\rangle$ 
in the MV model. 
We have expressed the number density of the produced gluons and 
their mean free path in terms of
$Q_s\tau_c$. Assuming $\tau_c=0.5$ fm/c with the saturation momentum, $Q_s\sim 2$ GeV,
we have shown that
the produced gluons can be thermalized within the period, $\tau_c$.
When the saturation momentum becomes larger, the thermalization
is achieved earlier.
In our discussion 
inhomogeneous color magnetic field has been assumed to decay
with the production of the
Nielsen-Olesen unstable modes. They are supposed to be
induced by an effective homogeneous color magnetic
field given by
the root mean square of the original color magnetic flux.
In order to see more precisely the decay,
we need to examine more closely how the inhomogeneous color magnetic field 
decays generating the unstable modes.

\vspace*{2em}
We would like to express thanks
to Dr. Itakura for useful comments.

%%%%%%%%%%%%%%%%%%%%%%%

\end{document}